\documentclass[reprint,showpacs,superscriptaddress,aps,nofootinbib]{revtex4-1}
\usepackage{amssymb}
\usepackage{amsmath,amssymb,graphicx,amsfonts}
\usepackage{graphicx}
\usepackage{dcolumn}
\usepackage{bm}
\usepackage{xcolor,color,colortbl}
\definecolor{Gray}{gray}{0.9}
\RequirePackage{mathptmx}      

\def\beq{\begin{equation}}
\def\eeq{\end{equation}}
\def\bea{\begin{eqnarray}}
\def\eea{\end{eqnarray}}

\newcommand{\overbar}[1]{\mkern 1.5mu\overline{\mkern-1.5mu#1\mkern-1.5mu}\mkern 1.5mu}

\usepackage[T1]{fontenc}
\usepackage{pstricks}
\usepackage{graphicx}
\usepackage{amsbsy}
\usepackage{bm}
\usepackage{color}
\usepackage{fancyhdr}
\usepackage{epsfig}
\usepackage[colorlinks=true,linkcolor=blue,citecolor=green,urlcolor=blue,filecolor=blue]{hyperref}
\usepackage{slashed}
\usepackage{multirow}
\usepackage{slashed}

\begin{document}

\bigskip

\vspace{2cm}
\title{Lepton number violation in $B_s$ meson decays induced by an on-shell Majorana neutrino}
\vskip 6ex

\author{Jhovanny Mej\'{i}a-Guisao}
\email{jmejia@fis.cinvestav.mx}
\affiliation{Departamento de F\'{i}sica, Centro de Investigaci\'{o}n y de Estudios Avanzados del IPN, Apartado Postal 14-740, 07000 Ciudad de M\'{e}xico, M\'{e}xico}
\author{Diego Milan\'{e}s}
\email{diego.milanes@cern.ch}
\affiliation{Departamento de F\'{i}sica, Universidad Nacional de Colombia, C\'{o}digo Postal 11001, Bogot\'{a}, Colombia}
\author{N\'{e}stor Quintero}
\email{nestor.quintero01@usc.edu.co}
\affiliation{Facultad de Ciencias B\'{a}sicas, Universidad Santiago de Cali, Campus Pampalinda, Calle
5 No. 62-00, C\'{o}digo Postal 76001, Santiago de Cali, Colombia}
\author{Jos\'{e} D. Ruiz-\'{A}lvarez}
\email{jose.ruiz@cern.ch}
\affiliation{Departamento de F\'{i}sica, Universidad de Los Andes, C\'{o}digo Postal 111711, Bogot\'{a}, Colombia}

\bigskip
\begin{abstract}
Lepton-number violation can be induced by the exchange of an on-shell Majorana neutrino $N$ in semileptonic $|\Delta L|=2$ decays of the $B_s$ meson, $B_s^0 \to P^-\pi^-\mu^+ \mu^+$ with $P=K, D_s$. We investigate the production of such a heavy sterile neutrino through these four-body $\mu^+\mu^+$ channels and explore the sensitivity that can be reached at the LHCb and CMS experiments.  For heavy neutrino lifetimes of $\tau_N$ = [1, 100, 1000] ps and integrated luminosities collected of 10 and 50 fb${}^{-1}$ at the LHCb and 30, 300, and 3000 fb${}^{-1}$ at the CMS, we find a significant sensitivity on branching fractions of the orders ${\rm BR}(B_s^0 \to K^- \pi^-\mu^+ \mu^+) \lesssim \mathcal{O}(10^{-9} - 10^{-8})$ and  ${\rm BR}(B_s^0 \to D_s^- \pi^-\mu^+ \mu^+) \lesssim \mathcal{O}(10^{-8} - 10^{-7})$. In the kinematically allowed mass ranges of $m_N \in [0.25,4.77]$ GeV and $m_N \in [0.25,3.29]$ GeV, respectively, we exclude regions on the parameter space ($m_N,|V_{\mu N}|^{2}$) associated with the heavy neutrino, which could slightly improve the limits from $B^- \to \pi^+\mu^-\mu^-$ (LHCb).
 \end{abstract} 

\maketitle
\bigskip

\section{Introduction}  \label{Intro}

To discriminate if the light neutrinos are Majorana or Dirac fermions (i.e. if neutrinos are their own antiparticles or not) is one of the most important puzzles in the Standard Model (SM)~\cite{deGouvea:2013}. To date, it is already well confirmed by a diversity of neutrino oscillation experiments (solar, atmospheric, reactors and accelerators)~\cite{PDG} that light neutrinos are massive particles; however, the responsible underlying mechanism remains unknown, and different new physics (NP) scenarios beyond the SM predict the neutrinos to be Dirac or Majorana massive fermions~\cite{NeutrinoReviews}. If neutrinos are Dirac massive particles, the total lepton number $L$ is a conserved quantity in the nature, while if neutrinos turn out to be Majorana massive particles, $L$ will be not  longer conserved and will be violated~\cite{deGouvea:2013}. The most remarkable searches of lepton-number violating (LNV) signals are by looking for processes with $|\Delta L|=2$, in which the possible existence of Majorana neutrinos can be tested~\cite{deGouvea:2013}.
 
The smoking-gun LNV signal is the neutrinoless double-$\beta$ ($0\nu\beta\beta$) decay~\cite{Rodejohann:2011,Gomez-Cadenas,Vissani:2015}. Searches of this rare nuclear transition have been pursued for several decades by different experiments and up to now no positive signal has been observed~\cite{Rodejohann:2011,Gomez-Cadenas,Vissani:2015}. Currently, the best limits on their half-lives have been obtained from the nuclei $^{76}{\rm Ge}$~\cite{GERDA} and $^{136}{\rm Xe}$~\cite{EXO-200,KamLAND-Zen}.
Aside from the $0\nu\beta\beta$ decay, low-energy studies of rare semileptonic processes in $|\Delta L|= 2$ decays of pseudoscalar mesons ($K, D, D_s, B, B_c$) and the $\tau$ lepton have been considered as complementary and alternative evidence to prove the Majorana nature of neutrinos ~\cite{Atre:2009,Kovalenko:2000,Ali:2001,Atre:2005,Kovalenko:2005,Helo:2011,Cvetic:2010,
Zhang:2011,Bao:2013,Wang:2014,Quintero:2016,Sinha:2016,Gribanov:2001,Quintero:2011,Quintero:2012b,
Quintero:2013,Dong:2013,Yuan:2013,Quintero:2012a,Dib:2012,Dib:2014,Yuan:2017,
Shuve:2016,Asaka:2016,Cvetic:2016,Cvetic:2017,Zamora-Saa:2016,Cvetic:CP}. 
Since these $|\Delta L|= 2$ decays  can be produced (and enhanced) via an intermediate on-shell Majorana neutrino $N$ with a mass in the range $\sim$[0.1,5.0] GeV, the phenomenology associated with such a heavy neutrino has been actively studied~\cite{Atre:2009,Atre:2005,Kovalenko:2005,Helo:2011,Cvetic:2010,
Zhang:2011,Bao:2013,Wang:2014,Quintero:2016,Sinha:2016,Gribanov:2001,Quintero:2011,Quintero:2012b,
Quintero:2013,Dong:2013,Yuan:2013,Quintero:2012a,Dib:2012,Dib:2014,Yuan:2017,Shuve:2016,Asaka:2016,
Cvetic:2016,Cvetic:2017,Zamora-Saa:2016,Cvetic:CP}. From the experimental side, upper limits on the branching fractions of various LNV processes have been set by different experiments such NA48/2, BABAR, Belle, LHCb, and E791  \cite{CERNNA48/2:2016,BABAR,BABAR:2014,LHCb:2012,LHCb:2013,LHCb:2014,Belle:2011,Belle:2013,E791}. See also the Particle Data Group~\cite{PDG}.

Focusing on the $b$-quark sector, recent attention has been paid to the four-body $|\Delta L|=2$ decays of $B$ and $B_c$ mesons: $\bar{B}^0 \to D^+\pi^+\mu^{-}\mu^{-}$, $B^{-} \to D^{0}\pi^{+}\mu^{-}\mu^{-}$~\cite{Quintero:2011,Quintero:2013,Cvetic:2016,Cvetic:2017}, and $B_c^- \to J/\psi\pi^+\mu^-\mu^-$~\cite{Quintero:2016,Sinha:2016}. In addition, the $|\Delta L|=2$ decays of $\Lambda_b$ baryon have been explored as well~\cite{Mejia-Guisao:2017}. As a salient feature, these decay channels are not highly suppressed by Cabbibo-Kobayashi-Maskawa (CKM) factors, and their experimental search is within reach of sensitivity of the LHCb and Belle II~\cite{Cvetic:2017,Quintero:2016,Sinha:2016}. So far, the LHCb has reported the upper limit ${\rm BR}(B^{-} \to D^{0} \pi^{+}\mu^{-}\mu^{-})$ $< 1.5\times 10^{-6}$~\cite{LHCb:2012}, and improvements are expected in Run 2 and the future upgrade Run 3. On the other hand, the same quark level LNV transition that generates these four-body $|\Delta L|=2$ channels, can also produce $|\Delta L|=2$ decays in the $B_s$ meson and their signals may be detected at the LHC, which offers an excellent environment for the $B_s$ physics.

In this work, we will explore the LNV decay channels of the $B_s$ meson, $B_s^0 \to P^-\pi^-\mu^+ \mu^+$ with $P=K, D_s$, via an intermediate GeV-scale on-shell Majorana neutrino $N$. To our knowledge, these $|\Delta L|=2$ decays have not been investigated before from a theoretical nor from an experimental point of view. 
We will work in a simplified approach in which one heavy neutrino $N$ mixes with one flavor of SM lepton $\ell$ and its interactions are completely determined by the mixing angle $V_{\ell N}$~\cite{Atre:2009}.
Since $0\nu\beta\beta$ decay puts stringent limits to the electron-heavy neutrino mixing $|V_{eN}|^{2} \lesssim 10^{-8}$~\cite{Kovalenko:2014}, we will focus our attention on the above four-body $\mu^+\mu^+$ modes and explore their expected sensitivities at the LHCb and CMS experiments. We will show that their experimental search allows us to scan the parameter space $(m_N,|V_{\mu N}|^2)$ of the heavy neutrino sector, therefore, an additional test of the existence of Majorana neutrinos. 

Let us mention that the presence of a heavy neutrino with a mass of few GeV [$\sim \mathcal{O}(1)$ GeV] provides a realistic and falsifiable scenario for a common explanation of the baryon asymmetry of the Universe via leptogenesis~\cite{Shaposhnikov:2005,Shaposhnikov:2013,Drewes:2014,GeV_Leptogenesis} and the generation of neutrino masses via the GeV-scale seesaw model~\cite{Rasmussen:2016,deGouvea:2007}. This give us further motivation to study $|\Delta L|=2$ decays of the $B_s$ meson under consideration.

This work is organized as follows. In Sec. \ref{LNV_Bs}, we study the $|\Delta L|=2$ decays of the $B_s$ meson. The expected experimental sensitivities for these channels at the LHCb and CMS is discussed in Sec. \ref{sensitivity}. Based on the results of the previous, in Sec. \ref{constraints}, we discuss the bounds on the parameter space  $(m_N,|V_{\mu N}|^2)$ of the heavy neutrino that can be achieved. Our conclusions are given in Sec. \ref{Conclusion}.

\section{LNV decays of $B_s$ meson}  \label{LNV_Bs}

In this section, we study LNV signals in the  $|\Delta L|= 2$ decays of the $B_s$ meson $B_s^0 \to P^- \pi^-\mu^+ \mu^+$, with $P = K, D_s$  denoting a final-state pseudoscalar meson. These processes can occur via intermediate on-shell Majorana neutrino $N$ through the semileptonic decay $B_s^0 \to P^- \mu^+N$ followed by the subsequent decay $N \to \mu^+\pi^-$, with a kinematically allowed mass in the ranges
\begin{eqnarray}
B_s^0 &\to & K^- \pi^-\mu^+ \mu^+ :  \ \ m_N \in [0.25,4.77] \ {\rm GeV}, \nonumber \\
B_s^0 &\to & D_s^-\pi^-\mu^+ \mu^+ :  \ \ m_N \in [0.25,3.29] \ {\rm GeV}. \nonumber
\end{eqnarray}

\noindent The $B_s^0 \to P^- \pi^-\mu^+ \mu^+$ decays are then split into two subprocesses and the corresponding branching fraction can be written in the factorized form 
\begin{eqnarray}\label{4leptonic}
{\rm BR}(B_s^0  \to P^-\pi^-\mu^+\mu^+) &=& {\rm BR}(B_s^0 \to P^- \mu^+ N) \nonumber \\
&& \times  \Gamma(N \to \mu^+\pi^-) \tau_N  /\hbar,
\end{eqnarray}  

\noindent  with $\tau_N$ as the lifetime of the Majorana neutrino. The branching ratio of $B_s^0 \to P^-  \mu^+ N$ is given by the expression~\cite{Cvetic:2016} 
\begin{equation}
{\rm BR}(B_s^0 \rightarrow P^-  \mu^+ N) =  |V_{\mu N}|^2  \int dt \dfrac{d\overbar{\rm BR}(B_s^0 \rightarrow P^-  \mu^+ N)}{dt} ,
\end{equation}

\noindent where
\begin{eqnarray} \label{BR_Bs}
&& \dfrac{d\overbar{\rm BR}(B_s^0 \rightarrow P^-  \mu^+ N)}{dt} =\nonumber  \\ 
&&  \dfrac{G_F^2 \tau_{B_s}}{384\pi^3 m_{B_s}^3 \hbar} |V_{qb}^{\rm CKM}|^2 \ \dfrac{\big[\lambda(m_\mu^2,m_N^2,t)\lambda(m_{B_s}^2,m_P^2,t) \big]^{1/2}}{t^3} \nonumber \\
&& \times  \Big(\big[F_+^{B_sP}(t)\big]^2 C_+(t) + \big[F_0^{B_sP}(t)\big]^2 C_0(t) \Big) ,
\end{eqnarray} 

\noindent is the so-called differential canonical branching ratio~\cite{Cvetic:2016}, where $G_F$ is the Fermi constant; $V_{qb}^{\text{CKM}}$ denotes the CKM matrix element involved (with $q = u, c$ for $P=K, D_s$)\footnote{We will use the central values $|V_{ub}^{\text{CKM}}|= 4.09 \times 10^{-3}$ and $|V_{cb}^{\rm{CKM}}| = 40.5 \times 10^{-3}$~\cite{PDG}.}; and $F_{+}^{B_sP}(t)$ and $F_{0}^{B_sP}(t)$ are the vector and scalar form factors for the $B_s \to P$ transition, respectively, which are eva\-lua\-ted at the square of the transferred momentum $t=(p_{B_s}-p_P)^2$.  The usual kinematic K\"{a}llen function is denoted by $\lambda(x,y,z)=x^{2}+y^{2}+z^{2}-2(xy+xz+yz)$. The coefficients $C_+(t)$ and $C_0(t)$ in~\eqref{BR_Bs} are defined as
\begin{eqnarray}
C_+(t) &=& \lambda(m_{B_s}^2,m_P^2,t) [2t^2 + t(m_\mu^2 + m_N^2) + (m_\mu^2 - m_N^2)^2],\nonumber\\ 
&&  \\
C_0(t) &=& 3(m_{B_s}^2 - m_P^2) [m_\mu^2(t+2m_N^2 -m_\mu^2) \nonumber\\ 
&&+ m_N^2(t -m_N^2)],
\end{eqnarray}

\noindent respectively. The total branching fraction is then obtained by integrating the differential canonical branching ratio over the full $t$ region $[(m_\mu+m_N)^2$,$(m_{B_s}-m_P)^2]$.

As mentioned at the Introduction, the coupling of the heavy neutrino (sterile) $N$ to the charged current of lepton flavor $\mu$ is characterized by the quantity $V_{\mu N}$~\cite{Atre:2009}. Without referring to any NP scenario, we will treat $m_N$ and $V_{\mu N}$ as unknown phenomenological parameters that can be constrained (set) from the experimental non-observation (observation) of $|\Delta L| =2$ processes~\cite{Atre:2009,Helo:2011,Quintero:2016}. 

On the other hand, the decay width of $N \to \mu^+\pi^-$ is given by the expre\-ssi\-on \cite{Atre:2009}
\begin{equation}
\Gamma(N  \to \mu^+\pi^-) = |V_{\mu N}|^2 \ \bar{\Gamma}(N  \to  \mu^-\pi^+),
\end{equation}

\noindent with
\begin{eqnarray}
\bar{\Gamma}(N \to \mu^+\pi^-) &=&  \dfrac{G_F^2}{16 \pi}|V_{ud}^{\text{CKM}}|^2  f_\pi^2 m_N \sqrt{\lambda(m_N^2,m_\mu^2,m_\pi^2)}\nonumber \\
&& \times \bigg[ \bigg(1- \dfrac{m_\mu^2}{m_N^2} \bigg)^2 - \dfrac{m_\pi^2}{m_N^2} \bigg(1+ \dfrac{m_\mu^2}{m_N^2} \bigg) \bigg], \nonumber \\ \label{Ntopimu}
\end{eqnarray}

\noindent where  $|V_{ud}^{\text{CKM}}|= 0.97417$~\cite{PDG} and $f_{\pi}= 130.2(1.7)$ MeV is the pion decay constant~\cite{Rosner:2015}. 

The lifetime of the Majorana neutrino $\tau_N=\hbar / \Gamma_N$ in Eq.~\eqref{4leptonic} can be obtained by summing over all accessible final states that can be opened at the mass $m_N$~\cite{Atre:2009}. However, in further analysis (Secs. \ref{sensitivity} and \ref{constraints}), we will leave it as a phenomenological parameter accessible to the LHCb and CMS experiments.

\subsection{Form factors $B_s \to P$ ($P=K,D_s$)}  \label{FF_BsP}

For the form factors associated with the $B_s \to P$ transition, we will use the theoretical predictions provided by the lattice QCD approach~\cite{Monahan:2017,Bouchard:2014}. 

The form factors $F_+^{B_s D_s}$ and $F_0^{B_s D_s}$ can be represented by the  $z$ expansion through a modification of the Bourrely-Caprini-Lellouch (BCL) parametrization~\cite{Monahan:2017},
\begin{eqnarray}
F_+^{B_sP}(t) &=& \dfrac{1}{(1-t/M_+^2)}\sum_{n=0}^{J-1} b_n^+ \Big[z(t)^n - (-1)^{n-J} \dfrac{n}{J}z(t)^J \Big], \label{F1_BsDs} \nonumber \\
\\
F_0^{B_sP}(t) &=& \dfrac{1}{(1-t/M_0^2)} \sum_{n=0}^{J-1} b_n^0 z(t)^n, \label{F0_BsDs}
\end{eqnarray}

\noindent respectively, where the $z(t)$ function is defined as
\begin{equation}
z(t) = \dfrac{\sqrt{t_+ - t}-\sqrt{t_+ - t_0}}{\sqrt{t_+ - t}+\sqrt{t_+ - t_0}}.
\end{equation}

\noindent In Table~\ref{Lattice}, we show the respective coefficients of the $z$ expansion in Eqs.~\eqref{F1_BsDs} and~\eqref{F0_BsDs} for $J=3$ as well as additional parameters: pole masses $M_{+(0)}$ and $t_{+(0)}$~\cite{Monahan:2017}. The masses of particles involved are taken from the Particle Data Group~\cite{PDG}.

\begin{table}[!t]
\centering
\renewcommand{\arraystretch}{1.2}
\renewcommand{\arrayrulewidth}{0.8pt}
\caption{\small Coefficients $(b_0^+,b_1^+,b_2^+)$ and $(b_0^0,b_1^0,b_2^0)$ of the $z$ expansion in Eqs. \eqref{F1_BsDs} and \eqref{F0_BsDs}, pole masses $M_{+(0)}$ and $t_{+(0)}$.}
\begin{tabular}{cc}
\hline
Parameter &$B_s \to D_s$~\cite{Monahan:2017}  \\
\hline
$M_+$ (GeV) & 6.330 \\
$M_0$ (GeV) & 6.420 \\
$t_+$ (GeV$^2$)   & $(m_{B_s}+m_{D_s})^2$ \\
$t_0$ (GeV$^2$) & $(m_{B_s}-m_{D_s})^2$\\
$b_0^+$& 0.858\\
$b_1^+$ & -3.38\\
$b_2^+$  & 0.6 \\
$b_0^0$  & 0.658\\
$b_1^0$  & -0.10\\
$b_2^0$ & 1.3 \\
\hline
\end{tabular} \label{Lattice}
\end{table}

In Ref.~\cite{Bouchard:2014}, the form factors for the $B_s \to K$ transition are parametrized in a modified BCL form
\begin{eqnarray}
F_+^{B_sK}(t) &=& \dfrac{1}{(1-t/M_+^2)} \sum_{n=0}^{2} a_n^+ \Big[z(t)^n - (-1)^{n-3} \dfrac{n}{3}z(t)^3 \Big],\label{F1_BsK} \nonumber \\
\\
F_0^{B_sK}(t) &=& \sum_{n=1}^{3} a_n^0 \big(z(t)^n -z(0)^n\big) \nonumber \\
&& \ + \sum_{n=0}^{2} a_n^+ \Big[z(0)^n - (-1)^{n-3} \dfrac{n}{3}z(0)^3\Big], \label{F0_BsK}
\end{eqnarray}

\noindent where the corresponding expansion coefficients $(a_0^+,a_1^+,a_2^+)$ and $(a_1^0,a_2^0,a_3^0)$, pole masses $M_{+(0)}$ and  $t_{+(0)}$, are displayed in Table~\ref{LatticeBsK}.

\begin{table}[!t]
\centering
\renewcommand{\arraystretch}{1.2}
\renewcommand{\arrayrulewidth}{0.8pt}
\caption{\small Coefficients $(a_0^+,a_1^+,a_2^+)$ and $(a_1^0,a_2^0,a_3^0)$ of the $z$-expansion in Eqs. \eqref{F1_BsK} and \eqref{F0_BsK}, pole masses $M_{+(0)}$ and $t_{+(0)}$.}
\begin{tabular}{cc}
\hline
Parameter & $B_s \to K$~\cite{Bouchard:2014} \\
\hline
$M_+$ (GeV) & 5.3252 \\
$M_0$ (GeV) & 5.6794  \\
$t_+$ (GeV$^2$) & $(m_{B_s}+m_K)^2$   \\
$t_0$ (GeV$^2$) &  $(m_{B_s}+m_K)(\sqrt{m_{B_s}}-\sqrt{m_K})^2$ \\
$a_0^+$ & 0.368 \\
$a_1^+$ & -0.750 \\
$a_2^+$ & 2.72\\
$a_1^0$ & 0.315\\
$a_2^0$ & 0.945 \\
$a_3^0$ & 2.391 \\
\hline
\end{tabular} \label{LatticeBsK}
\end{table}

\section{Expected experimental sensitivity at the LHC}  \label{sensitivity}

Now, let us provide an estimation of the expected number of events at the LHC, namely, LHCb and CMS experiments, for the $|\Delta L|=2$ channels of the $B_s$ meson, $B_{s}^0 \to P^-\pi^-\mu^+\mu^+$ (with $P = K, D_s$), discussed above.

\subsection{LHCb experiment} \label{LHCb}

The number of expected events in the LHCb experiment has the form 
\begin{eqnarray} \label{N:LHCb}
N_{\rm exp}^{\rm LHCb} &=&  \sigma(pp\to H_b X)_{\rm acc}f(b\to B_s) {\rm BR}(B_s \to \Delta L=2)\nonumber \\
&& \times  \epsilon_D^{\rm LHCb}(B_s \to \Delta L=2) P_N^{\rm LHCb} \ \mathcal{L}^{\rm LHCb}_{\rm int},
\end{eqnarray}

\noindent where $\sigma(pp\to H_b X)_{\rm acc}$ is the production cross section of $b$-hadrons inside the LHCb geometrical acceptance; $f(b\to B_s)$ is the hadronization factor of a $b$-quark to the $B_s$ meson; $\mathcal{L}_{\rm int}^{\rm LHCb}$ is the integrated luminosity; ${\rm BR}(B_s \to \Delta L=2)$ corresponds to the branching fraction of the given LNV process and $\epsilon_D^{\rm LHCb}(B_s \to \Delta L=2)$ is its detection efficiency of the LHCb detector involving  reconstruction, selection, trigger, particle misidentification, and detection efficiencies. Most of the the on-shell neutrinos produced in the decays  $B_{s}^0 \to (K^-,D_s^-)\mu^+N$ are expected to live a long enough time to travel through the detector and decay ($N \to  \mu^+ \pi^-$) far from the interaction region. This effect is given by the $P_N^{\rm LHCb}$ factor (acceptance factor), which accounts for the probability of the on-shell neutrino $N$ decay products to be inside the LHCb detector acceptance \cite{Dib:2014}. The reconstruction efficiency will depend on this acceptance factor as well.

The production cross section has been measured to be $\sigma(pp\to H_b X)_{\rm acc}=(75.3 \pm 5.4 \pm  13.0)$ $\mu$b inside the LHCb acceptance~\cite{Aaij:2010gn}. The world average for the hadronization factor is taken to be $f(b\to B_s)=0.103\pm0.005$~\cite{Amhis:2016}. The proper computation of the detection efficiency requires fully simulated Monte Carlo samples of the exclusive decay, reconstructed in the same way as real LHCb data.  Here, we perform a rough estimation of the detection efficiency, based on extrapolation of detection efficiencies already reported by LHCb experiment of similar final states.

\begin{figure}[!t]
\centering
\includegraphics[scale=0.445]{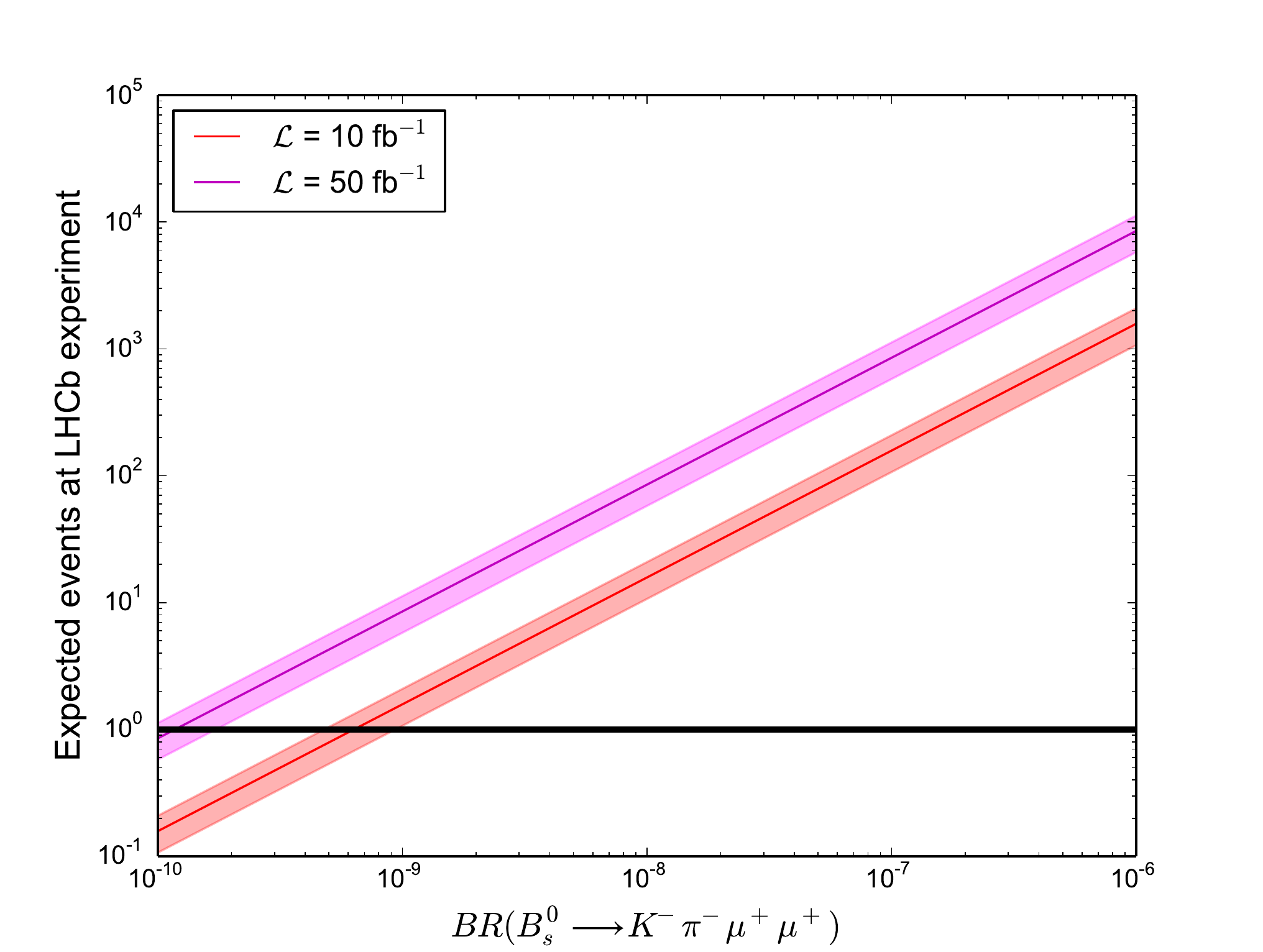} 
\includegraphics[scale=0.445]{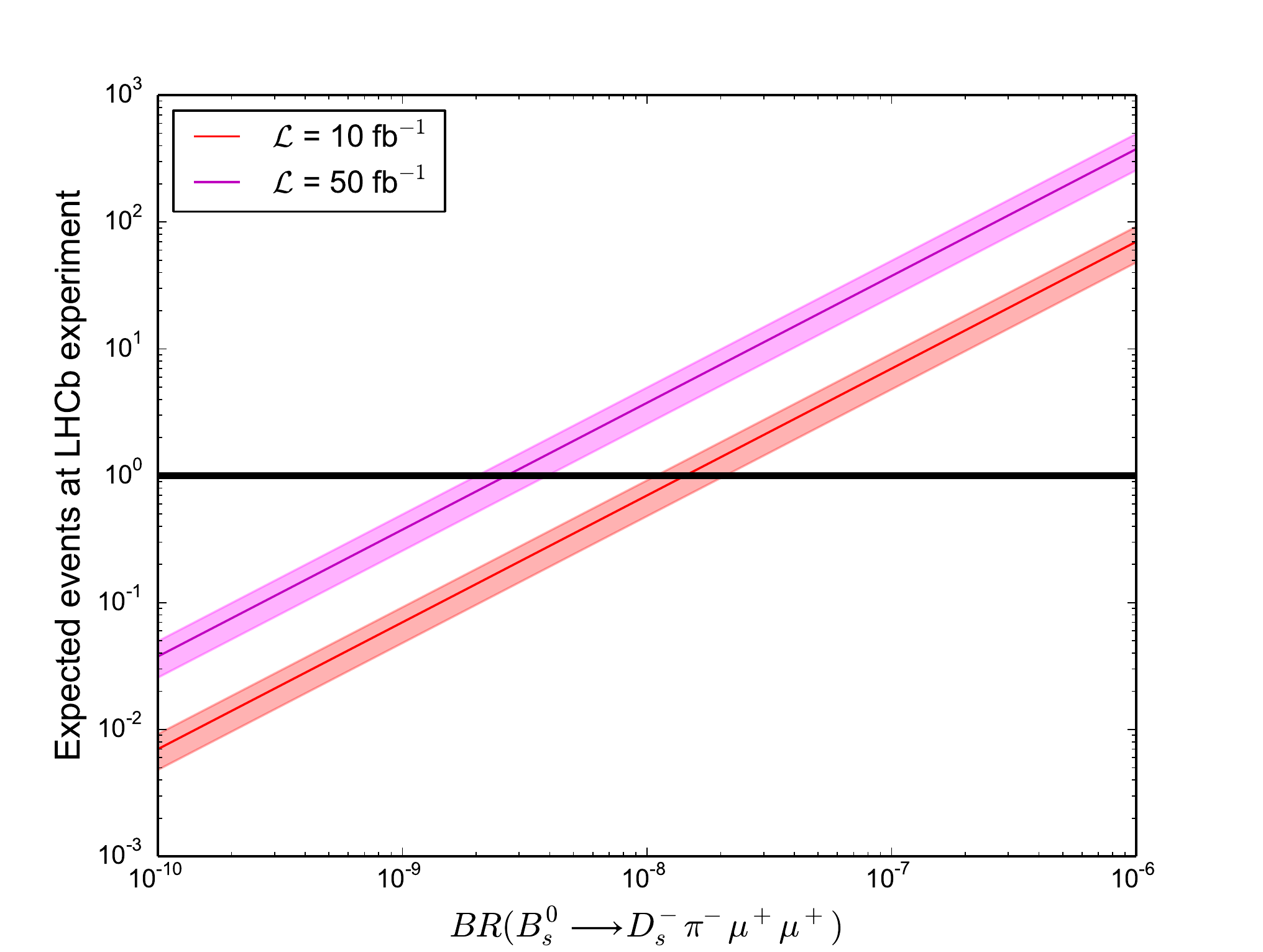}
\caption{\small  Number of expected events of the process $B_{s}^0 \to K^-\pi^-\mu^+\mu^+$ (top) and $B_{s}^0 \to D_s^-\pi^-\mu^+\mu^+$ (bottom) to be observed in the LHCb experiment as a function of their branching fractions for a luminosity of 10 fb$^{-1}$ (red) and 50 fb$^{-1}$ (magenta). The solid black line shows the central value, while the filled area shows the 1-$\sigma$ uncertainty.}
\label{Fig:1} 
\end{figure}

The LHCb Collaboration has measured the detection efficiency of the $B_s^0\to\phi(K^+K^-)\mu^+\mu^-$ decay mode to be $1.1\%$ ~\cite{Aaij:2015esa}. This measurement includes trigger, tracking, reconstruction, particle identification, and selection efficiency. Given the content of final-state charged tracks, we can consider the $B_{s}^0 \to K^-\pi^-\mu^+\mu^+$ to be the same as for the $B_s\to\phi(K^+K^-)\mu^+\mu^-$ decay. Regarding the $B_{s}^0 \to D_s^-\pi^-\mu^+\mu^+$ decay, a golden mode to reconstruct the $D_s^+$ meson hadronically is $D_s^+\to K^+K^-\pi^+$, where ${\rm BR}(D_s^+\to K^+K^-\pi^+)=(5.45\pm 0.17)\times 10^{-2}$~\cite{PDG}. In this situation, there will be two additional charged tracks in the final state; thus, we can multiply previous efficiency by 0.9 for each additional charged track, the approximated single track reconstruction efficiency at LHCb. Finally, in Ref.~\cite{Aaij:2016xmb}, reconstruction efficiencies for hypothetical long-lived particles inside the LHCb acceptance are given. Here, we can observe that a maximum variation of about 25\% is measured in the efficiencies of particles living in the [5 - 50] ps range, with masses up to 200 GeV; however, in our case, long-lived particles can only be produced on-shell, therefore with masses around few GeV.  Thus, to account for this effect, we will just add a 25\% relative uncertainty to our efficiency prediction, obtaining finally
\begin{eqnarray}
\epsilon_D^{\rm LHCb}(B_s\to K^-\pi^-\mu^+\mu^+) P_N^{\rm LHCb} &\simeq & (1.10\pm 0.27) \%, \nonumber \\ \nonumber\label{Fig:2} 
\epsilon_D^{\rm LHCb}(B_s\to D_s^-\pi^-\mu^+\mu^+)P_N^{\rm LHCb}   &\simeq & (0.89\pm 0.22) \% .\nonumber
\end{eqnarray}

\noindent With these values, the relative uncertainty on $N_{\text{exp}}^{\rm LHCb} $ is of 32\% for both LNV modes.

The LHCb experiment performance during LHC-Run1 can be found in Ref.~\cite{LHCb:performance}. During  LHC-Run2 the expectation is to collect 10\,fb$^{-1}$ at the LHC nominal construction energy of a center of mass of 14\,TeV. Already some work has been developed for the future LHCb upgrade, LHC-Run3, for which integrated luminosity of the order of 50 fb$^{-1}$ is expected. 
Assuming the above assumptions on efficiency and cross section, Fig.~\ref{Fig:1} shows the number of expected events to be observed in the LHCb experiment as a function of branching fraction for $|\Delta L| =2$ modes of $B_s$ meson. The figure shows  red and magenta functions, corresponding to LHC-Run2 and LHC-Run3, respectively. Table~\ref{BR:LHCb} shows the expected signal events at the LHCb experiment for some selected values of the branching ratio, given LHC-Run2 and LHC-Run3 expected integrated luminosities.
We can see that values of the branching fractions of the order $\mathcal{O}(10^{-9} - 10^{-8})$ for $B_s^0\to K^-\pi^-\mu^+\mu^+$ and $\mathcal{O}(10^{-8} - 10^{-7})$ for $B_s^0\to D_s^-\pi^-\mu^+\mu^+$ might be within the experimental sensitivity of the LHCb. 

\begin{table}[!t]
\centering
\renewcommand{\arraystretch}{1.2}
\renewcommand{\arrayrulewidth}{0.8pt}
\caption{\small Number of expected events at the LHCb for some selected values of the branching ratio (BR) of $B_s^0\to K^-\pi^-\mu^+\mu^+$ and $B_s^0\to D_s^-\pi^-\mu^+\mu^+$.}
\begin{tabular}{lccc}
\hline
Mode & $\mathcal{L}_{\rm int}^{\rm LHCb}$ (fb$^{-1}$) & BR & $N_{\text{exp}}^{\rm LHCb}$ \\
\hline
 $B_s^0\to K^-\pi^-\mu^+\mu^+$ &  50 & $10^{-6}$ & $8522\pm2727$ \\
 & & $10^{-7}$ & $852\pm273$ \\
 & & $10^{-8}$ & $85\pm27$ \\
 & & $10^{-9}$ & $9\pm3$ \\
\cline{2-4}
 &  10 & $10^{-6}$ & $1583\pm506$ \\
 & & $10^{-7}$ & $158\pm51$ \\
 & & $10^{-8}$ & $16\pm5$ \\
 & & $10^{-9}$ & $2\pm1$ \\
  \hline
$B_s^0\to D_s^-\pi^-\mu^+\mu^+$ &  50 & $10^{-6}$ & $376\pm120$ \\
 & & $10^{-7}$ & $37\pm12$ \\
 & & $10^{-8}$ & $4\pm1$ \\
\cline{2-4}
 &  10 & $10^{-6}$ & $70\pm22$ \\
 & & $10^{-7}$ & $7\pm2$ \\
\hline
\end{tabular} \label{BR:LHCb}
\end{table}

 \subsection{CMS experiment} \label{CMS}

We also consider the possible sensitivity of the CMS experiment to the LNV signals from $B_{s}$ meson decays. The expected number of event for the CMS experiment is written as
\begin{eqnarray} \label{N:CMS}
N_{\rm exp}^{\rm CMS} &=&   \sigma(pp\to B_{s} X)  {\rm BR}(B_{s}\to \Delta L=2) \nonumber \\
&& \times \epsilon_D^{\rm CMS}(B_{s} \to \Delta L=2) P_N^{\rm CMS} \mathcal{L}_{\rm int}^{\rm CMS},
\end{eqnarray}

\noindent where $\mathcal{L}_{\rm int}^{\rm CMS}$ is the integrated luminosity recorded by the CMS experiment from proton-proton collisions delivered by the LHC; $\sigma(pp\to B_{s} X)$ is the $B_{s}$ meson production cross section in the CMS experiment acceptance; $\epsilon_{D}^{\rm CMS}(B_{s} \to \Delta L=2)$ is the efficiency to reconstruct and identify the signal events, which includes the trigger efficiency; $P_N^{\rm CMS}$ is a factor that accounts for the CMS experiment acceptance to the decay of the neutrino; and BR$(B_{s} \to \Delta L=2)$ is the $B_{s}$ meson branching fraction.

The CMS experiment acceptance to the signal depends on its tracker capabilities to reconstruct charged particles, especially pions and muons. Muons are reconstructed using the tracking system and the muon chambers, while pions are reconstructed by the tracker solely. The decay products from the $B_{s}$ are not very energetic. For this study, we consider that the muons and pions from signal events have a $p_{T}<20$ GeV. The CMS experiment has shown to be $90\%$ efficient in reconstructing of charged tracks in the mentioned $p_{T}$ range~\cite{Chatrchyan:2014fea}. However, these studies were performed for a center of mass energy of 7 TeV; we consider that these results also stand for 13 TeV. In addition, we also assume that the reconstruction efficiency of muons is $90\%$, following the results from Ref.~\cite{MuonEff}.

We use the same techniques as in Ref.~\cite{Mejia-Guisao:2017} to make a rough estimate of the CMS experiment efficiency to reconstruct the signal events. From some analyses performed with the CMS experiment for similar events~\cite{Chatrchyan:2011vh,Khachatryan:2015isa}, we can assume that the efficiency for the events from $B_{s}^0 \to K^{-}\pi^{-}\mu^{+}\mu^{+}$ will be approximately the same $(1.56 \pm 0.05)$\%. For the decay channel $B_{s}^0 \to D_{s}^{-}\pi^{-}\mu^{+}\mu^{+}$, we need to consider the further decay of the $D_{s}^-$ meson. With the CMS experiment, it is not possible to distinguish from a charged track left in the detector by a pion or a kaon. Therefore, we consider all the possible decays of $D_{s}^{-}$ into three charged tracks. Considering world averages for $K\pi\pi$ or $KK\pi$ decay branching fractions~\cite{PDG}, we can derive that the ${\rm BR}(D_{s}^{-} \to 3$ charged tracks) = $13.00 \pm 1.96$.
Taking into account this additional branching fraction and the fact that we need to identify two additional charged tracks, we can plug an additional $90\%$ efficiency factor for the track to obtain the total efficiency for the $B_{s}^0 \to D_{s}^{-}\pi^{-}\mu^{+}\mu^{+}$ channel. We obtain that $\epsilon_D^{\rm CMS}(B_{s}^0 \to D_{s}^{-}\pi^{-}\mu^{+}\mu^{+})$ = $(1.26 \pm 0.04)\%$.

Considering the distance the neutrino can fly in the detector, we restrict the discussion to lifetimes between $\tau_{N} = 1$ and 1000 ps, where the detector has sensitivity. The neutrino originates from the decay of $B_s$. The mean lifetime of $B_s$ meson is 1.505 ps~\cite{PDG}. Considering that the mean momentum of $B_s$ is 20 GeV, from Table 1 in Ref.~\cite{Chatrchyan:2011vh}, the Lorentz time dilation factor for $B_s$ is $\frac{p}{M}\approx 4$, implying a decay length of 0.2 cm. For the neutrino, we consider that $\frac{p}{M}\approx 1$ as it proceeds from the $B_s$ decay. Therefore, the total decay length of the neutrino, taking into account the initial decay length of $B_s$, is $L_{N} = 0.2$ cm (30.2 cm) for $\tau_{N} = 1$ ps (1000 ps) lifetime. Accordingly, with the studies performed in Ref.~\cite{Chatrchyan:2014fea}, the reconstruction efficiency in the tracker is degraded in terms of the distance in the tracker system from where the traces originate. From the same study, the reconstruction efficiency of tracks originating at 30 cm from the collision point is 55\%, while for just 1 cm, it is 100\%. The relative uncertainty applied on the overall reconstruction efficiency from CMS results is 18\%. It can be expected that differences from these assumptions would be found if a full study were done using the most recent energies used by the LHC. However, we expect to cover these differences by the uncertainty assigned.

Additionally, the cross section of the $B_{s}^{0}$ meson production from proton-proton collisions in the geometrical acceptance of the CMS experiment is obtained from~\cite{Chatrchyan:2011vh}. The $\sigma(pp \to B_s^{0})  \times {\rm BR}(B_{s}^{0}\to J/\psi \phi) = 6.9 \pm 0.6 \pm 0.6 $ nb at 7 TeV, and taking  BR$(B_{s}^{0}\to J/\psi  \phi) = (1.07 \pm 0.08)\times 10^{-3}$, the pure production cross section for proton-proton collisions at 7 TeV is $\sigma(pp \to B_{s}^{0})  = (6.45 \pm 0.09)\times 10^{3} $ nb. Thus, assuming that the cross section increases as the center of mass energy, the $B_{s}^{0}$ production cross section at 13 TeV proton-proton collisions is $\sigma(pp \to B_{s}^{0})  = (11.98 \pm 0.17) \times 10^{3} $ nb. 

\begin{figure}[!t]
\centering
\includegraphics[scale=0.445]{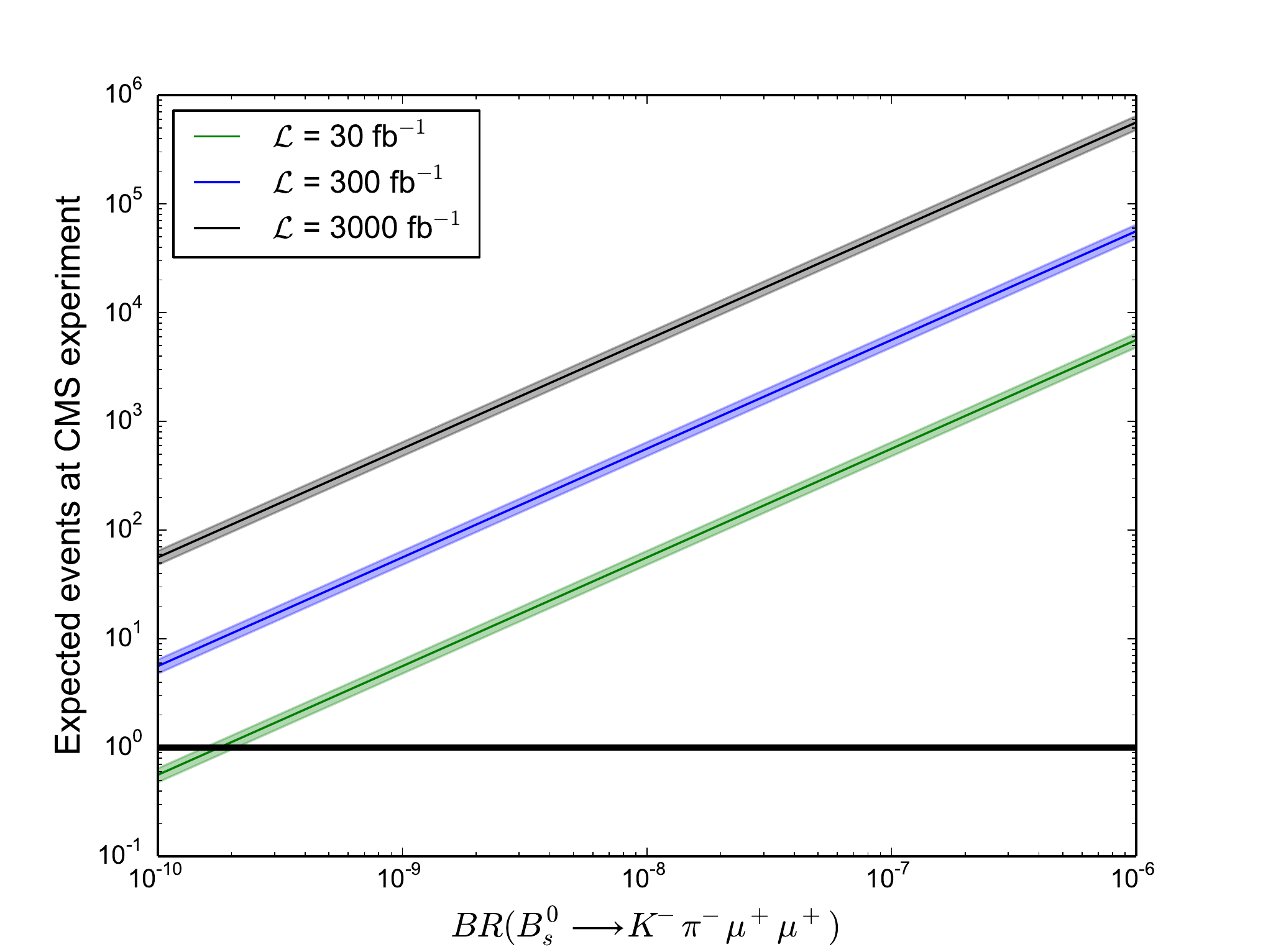} 
\includegraphics[scale=0.445]{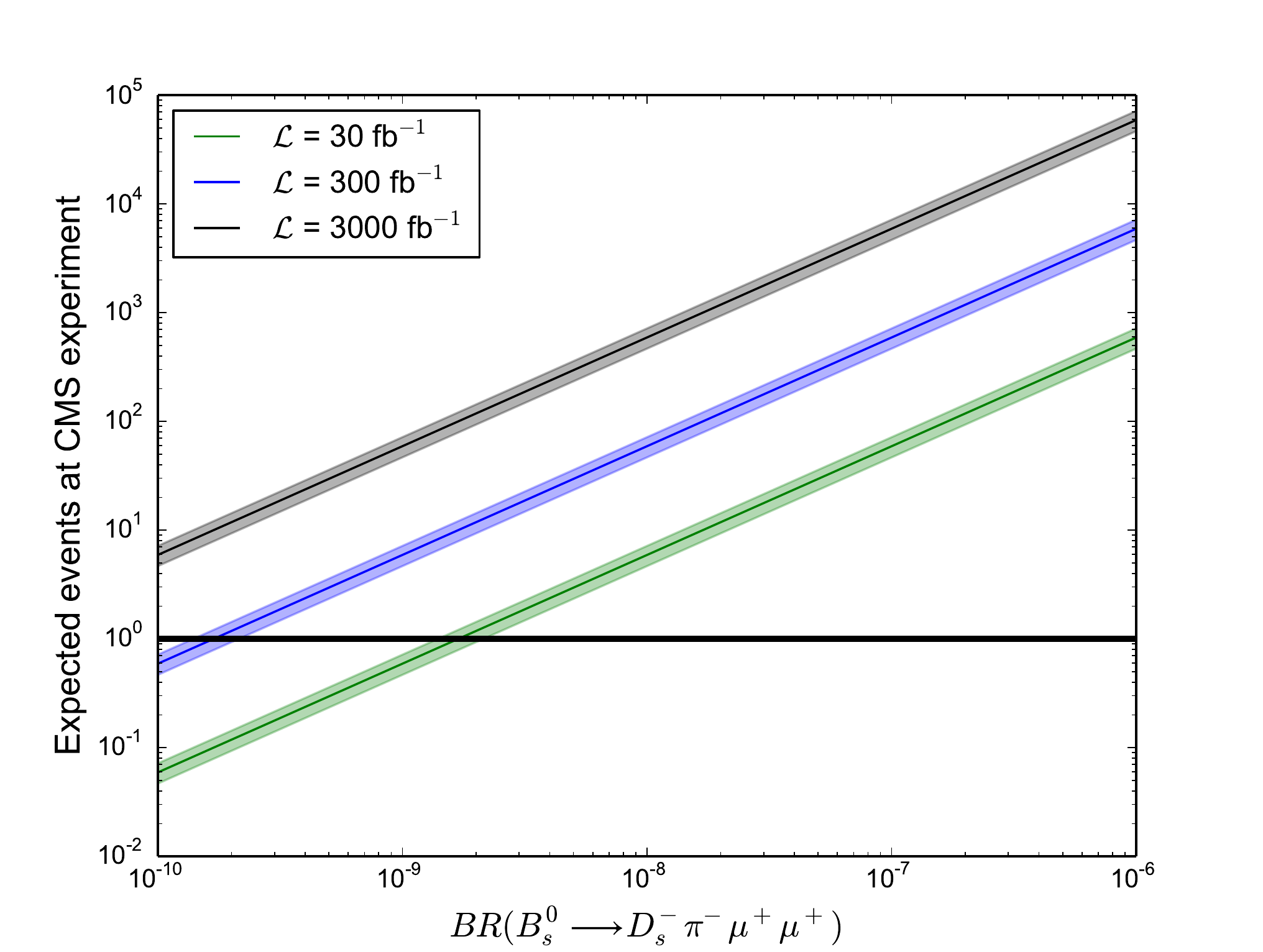}
\caption{\small  Expected events in the CMS experiment for $B_{s}^0 \to K^{-}\pi^{-}\mu^{+}\mu^{+}$ (top) and $B_{s}^0 \to D_{s}^{-}\pi^{-}\mu^{+}\mu^{+}$ (bottom) as a function of the branching fraction of the final state considered and for three benchmark luminosities: 30 (green), 300 (blue), and 3000 (gray) fb$^{-1}$. The central value is shown with a solid line. The shaded area represents the associated uncertainty in a 1-$\sigma$ window.}
\label{Fig:2} 
\end{figure}

Figure~\ref{Fig:2} shows the results for the expected number of events in the CMS experiment, using the above estimations. Three benchmark luminosities are used: $\mathcal{L}_{\rm int}^{\rm CMS}=$ 30, 300, and 3000 fb$^{-1}$. Table~\ref{BR:CMS} is used to quote explicitly some of the results obtained. We observe that for 30 and 300 fb$^{-1}$ the CMS experiment has sensitivity to branching fractions of the order  $\mathcal{O}(10^{-9} - 10^{-8})$ for $B_s^0\to K^-\pi^-\mu^+\mu^+$ and $\mathcal{O}(10^{-8} - 10^{-7})$ for $B_s^0\to D_s^-\pi^-\mu^+\mu^+$. Such a sensitivity is very similar to the one that can be reached by the LHCb (see Sec. \ref{LHCb}). We will consider these values of branching fractions as the most conservative ones to derive limits over the parameters of the heavy sterile neutrino in the next section.

\begin{table}[!t]
\centering
\renewcommand{\arraystretch}{1.2}
\renewcommand{\arrayrulewidth}{0.8pt}
\caption{\small Expected number of events for the CMS experiment with three branching fractions of 10$^{-6}$, 10$^{-7}$, and 10$^{-8}$ for $B_{s}^0 \to K^{-}\pi^{-}\mu^{+}\mu^{+}$ and $B_{s}^0 \to D_{s}^{-}\pi^{-}\mu^{+}\mu^{+}$.}
\begin{tabular}{lccc}
\hline
Mode & $\mathcal{L}_{\rm int}^{\rm CMS}$ (fb$^{-1}$) & BR & $N_{\text{exp}}^{\rm CMS}$ \\
\hline
 $B_{s}^0 \to K^{-}\pi^{-}\mu^{+}\mu^{+}$ & 30  & $10^{-6}$ & $5616 \pm 825$\\
 &  & $10^{-7}$ & $562 \pm 82$\\
 &   & $10^{-8}$ & $56 \pm 8$\\
\cline{2-4}
 & 300 & $10^{-8}$ & $562 \pm 82$ \\ 
  &  & $10^{-9}$ & $56 \pm 8$ \\
 \hline
$B_{s}^0 \to D_{s}^{-}\pi^{-}\mu^{+}\mu^{+}$ & 30  & $10^{-6}$  & $591 \pm 87$\\
 &  & $10^{-7}$  & $59 \pm 9$\\
  &   & $10^{-8}$  & $6 \pm 1$\\
  \cline{2-4}
   &  300  & $10^{-7}$ & $591 \pm 87$ \\ 
  &  & $10^{-8}$ & $59 \pm 9$ \\ 
\hline
\end{tabular} \label{BR:CMS}
\end{table}

\section{Bounds on the parameter space $(m_N,|V_{\mu N}|^2)$}  \label{constraints}

The experimental non-observation of $|\Delta L| =2$ processes can be  reinterpreted as bounds on the parameter space of a heavy sterile neutrino $(m_N,|V_{\mu N}|^2)$, namely, the squared mixing element $|V_{\mu N}|^2$ as a function of the mass  $m_N$~\cite{Atre:2009,Helo:2011,Quintero:2016}. Based on the analysis presented in Sec. \ref{sensitivity}, here, we explore the constraints on the $(m_N,|V_{\mu N}|^2)$ plane that can be achieved from the experimental searches on $B_s^0 \to  (K^-, D_s^-)$ $\pi^-\mu^+ \mu^+$ at the LHC, namely the LHCb and CMS experiments. 

From Eq.~\eqref{4leptonic}, it is straightforward to obtain the relation
\begin{equation}
|V_{\mu N}|^2 = \Bigg[  \frac{\hbar \ {\rm BR}(B_s^0 \to  P^- \pi^-\mu^+ \mu^+)}{ \overbar{\rm BR}(B_s^0 \to  P^- \mu^+ N) \times \overbar{\Gamma}(N \to \mu^+\pi^-) \tau_N}\Bigg]^{1/2} ,
\end{equation}

\noindent where $\overbar{{\rm BR}}(B_s^0 \to  P^-\mu^+N)$ and $\overbar{\Gamma}(N \to \mu^+\pi^-)$ are given by Eqs.~\eqref{BR_Bs} and~\eqref{Ntopimu}, respectively. As was already discussed in Sec.~\ref{sensitivity} and following the analysis of NA48/2~\cite{CERNNA48/2:2016} and LHCb~\cite{BABAR:2014}, we will consider heavy neutrino lifetimes of $\tau_N = [1, 100, 1000]$ ps as benchmark points in our analysis. This will allow us to extract limits on $|V_{\mu N}|^2$ without any additional assumption on the relative size of the mixing matrix elements.

\begin{figure}[!b]
\centering
\includegraphics[scale=0.43]{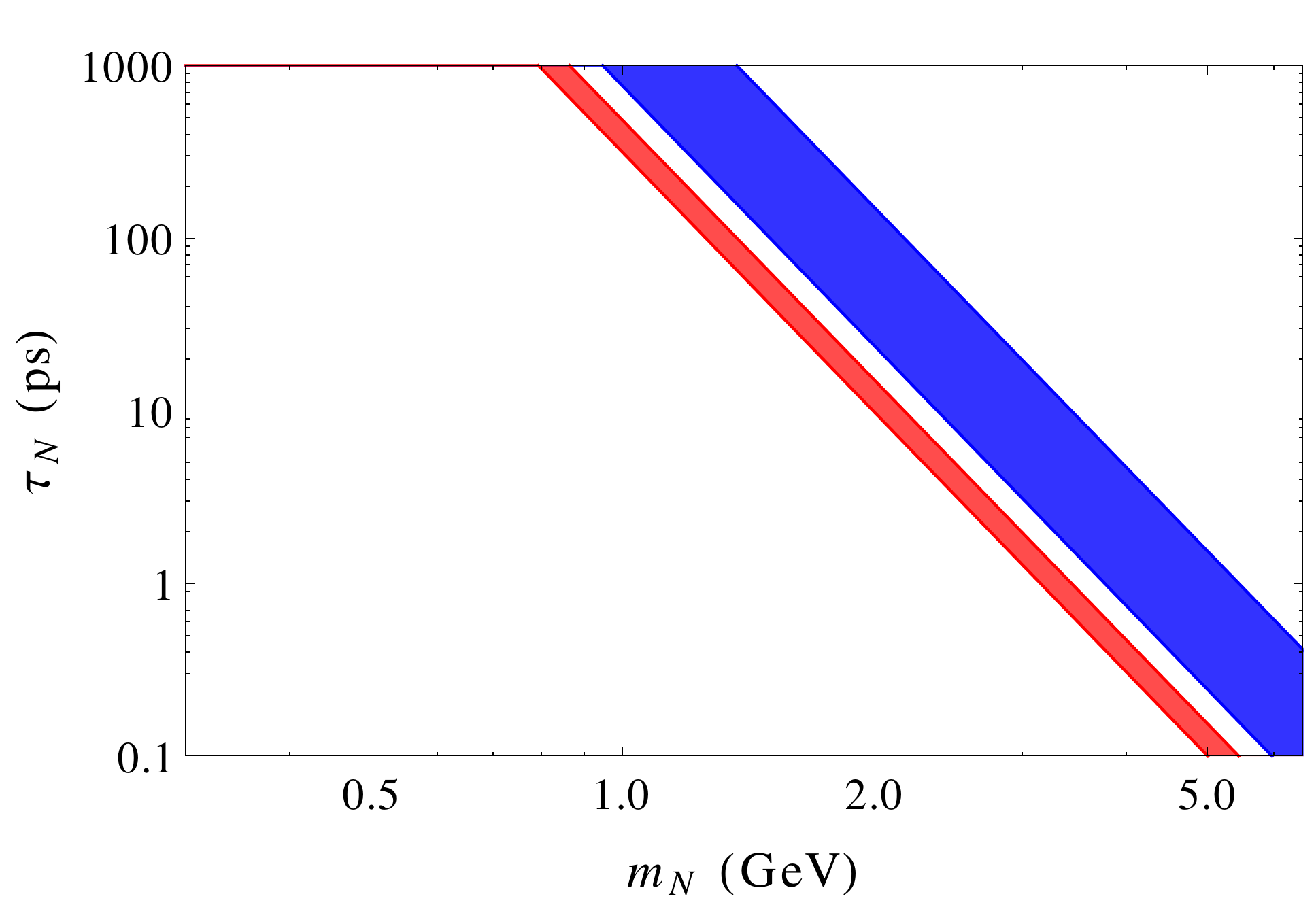}
\caption{\small Heavy neutrino lifetime $\tau_N$ as a function of $m_N$. The blue and red bands correspond to the allowed parameter space for $|V_{\tau N}|^{2}=10^{-3}$ and $10^{-2}$, respectively, while $|V_{eN}|^{2}$ and $|V_{\mu N}|^{2}$ vary within the range $[10^{-7},10^{-3}]$.}
\label{Fig4.0}
\end{figure}

From the theoretical point of view, is worth it to justify heavy neutrino lifetimes within the domain $1 \ {\rm ps}\leq \tau_N \leq 1000 \ {\rm ps}$ accessible to the LHCb and CMS experiments (see Secs.~\ref{LHCb} and~\ref{CMS}). For that purpose, we will use the approximate expression for the neutrino decay width
\begin{equation} \label{GammaN}
\Gamma_N = \dfrac{G_F m_N^5}{96\pi^3} \big[8 \big(|V_{eN}|^{2}+|V_{\mu N}|^{2} \big) +  3|V_{\tau N}|^{2} \big],
\end{equation}

\noindent which has been previously considered in the literature ~\cite{Cvetic:2016,Cvetic:2017,Cvetic:CP} for neutrino masses relevant to the $B_s$ meson decays under consideration. By considering the current bounds on 
$|V_{\ell N}|^{2}$ ($\ell = e, \mu, \tau$) given in Ref.~\cite{Atre:2009}, we will vary $|V_{eN}|^{2}$ and $|V_{\mu N}|^{2}$  within the range $[10^{-7},10^{-3}]$ and  $|V_{\tau N}|^{2}$ from $10^{-3}$ to $10^{-2}$. In Fig.~\ref{Fig4.0}, we plot the heavy neutrino lifetime $\tau_N = \hbar/\Gamma_N$ as a function of $m_N$. The blue and red bands correspond to the allowed parameter space for $|V_{\tau N}|^{2}=10^{-3}$ and $10^{-2}$, respectively.  According to Fig.~\ref{Fig4.0}, it is possible to obtain masses at the GeV-scale within the lifetimes domains accessible to the LHCb and CMS experiments.

\begin{figure}[!t]
\centering
\includegraphics[scale=0.43]{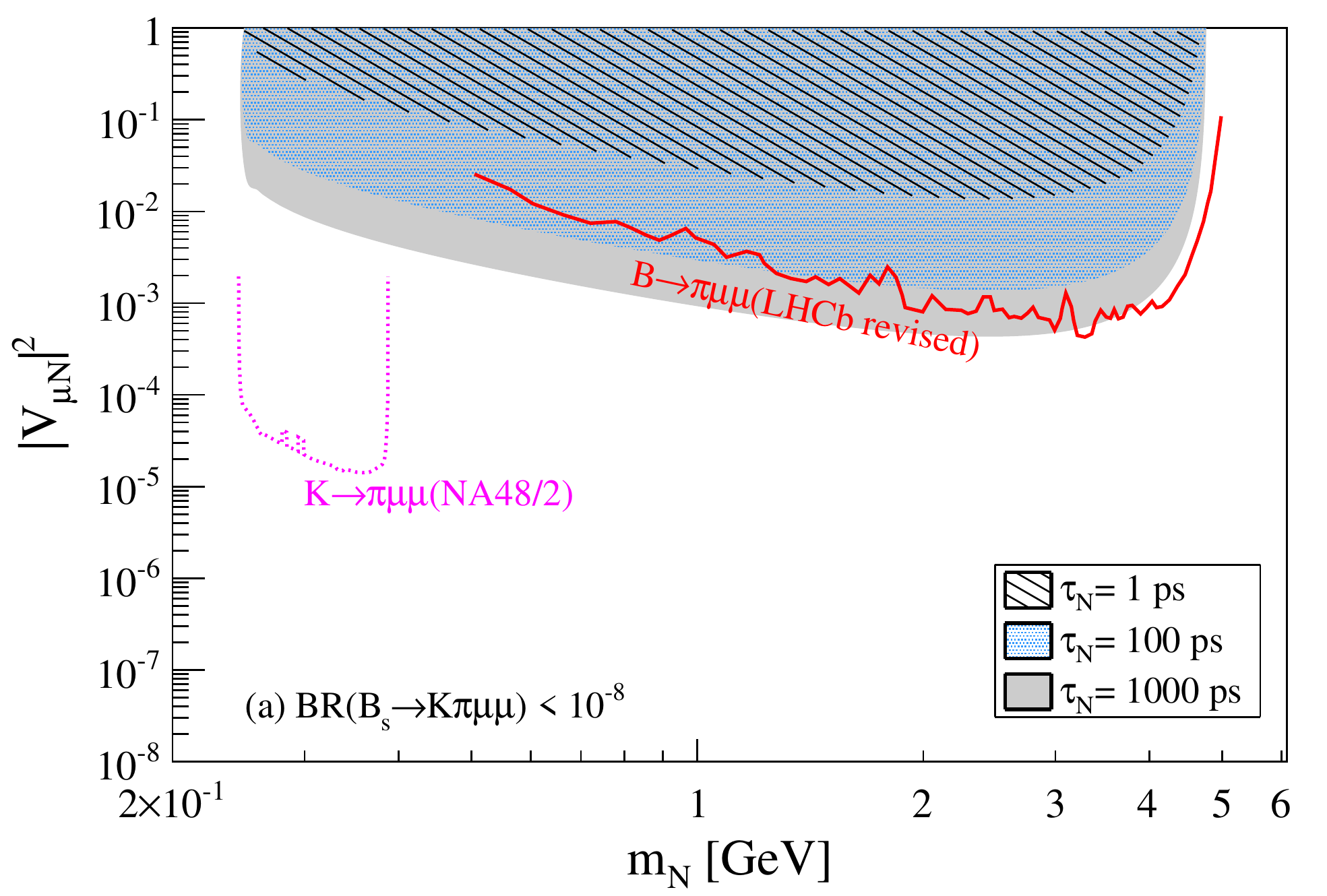}
\includegraphics[scale=0.43]{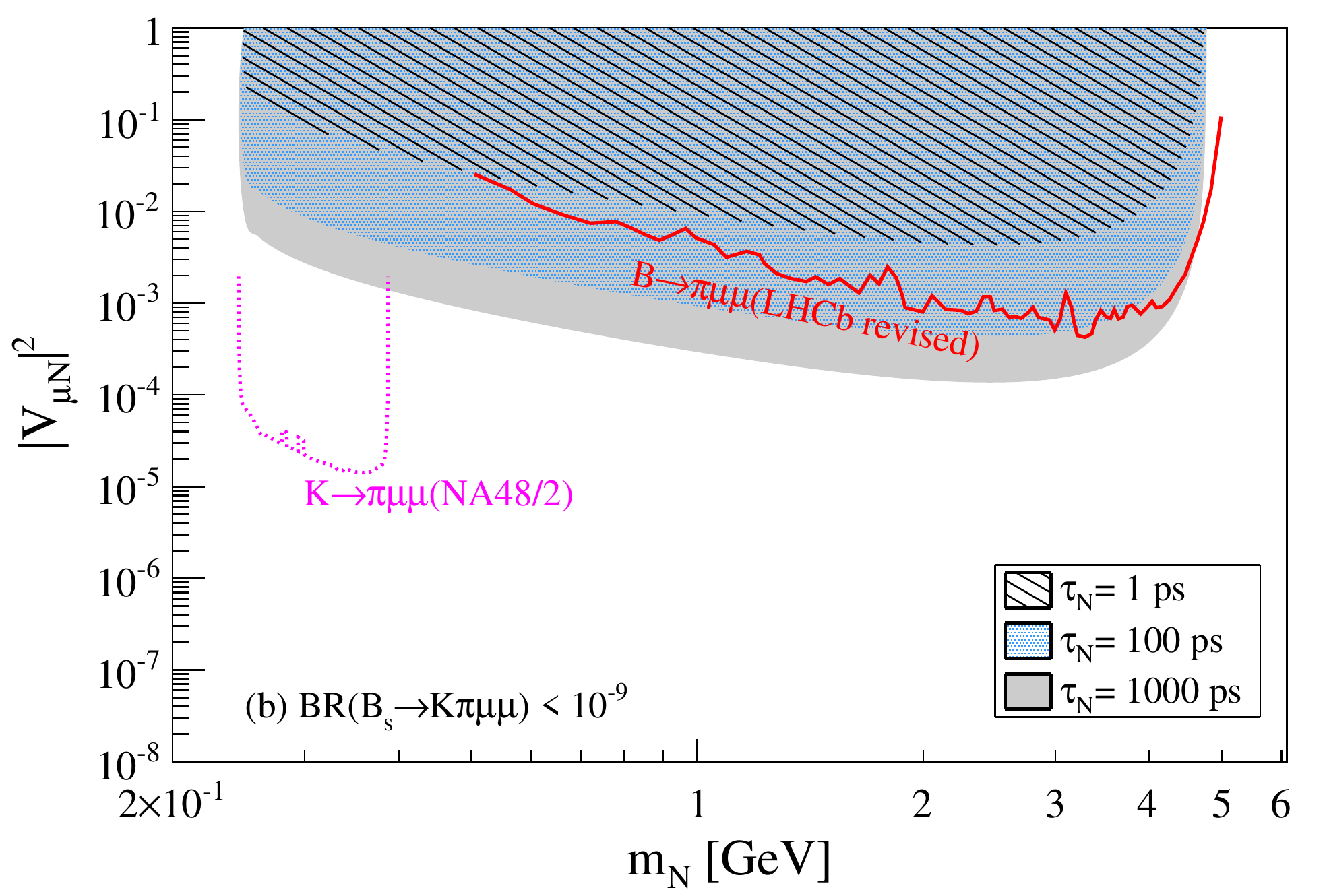}
\caption{\small Exclusion regions on the $(m_N, |V_{\mu N}|^{2})$ plane for (a) ${\rm BR}(B_s^0 \to  K^- \pi^-\mu^+ \mu^+) < 10^{-8}$ and  (b) ${\rm BR}(B_s^0 \to  K^- \pi^-\mu^+ \mu^+) < 10^{-9}$. The black, blue, and gray regions represent the bounds obtained for heavy neutrino lifetimes of $\tau_N = 1, 100, 1000$ ps, respectively. Limits provided by $K^- \rightarrow \pi^{+}\mu^{-}\mu^{-}$~\cite{CERNNA48/2:2016} and $B^{-} \to \pi^{+}\mu^{-}\mu^{-}$~\cite{Shuve:2016} are also included for comparison.}
\label{Fig4.1}
\end{figure}

In Figs.~\ref{Fig4.1}(a) and~\ref{Fig4.1}(b) we show the exclusions regions on $|V_{\mu N}|^{2}$ as a function of $m_N$ obtained by taking an expected sensitivity on the branching fractions of the orders ${\rm BR}(B_s^0 \to  K^- \pi^-\mu^+ \mu^+) < 10^{-8}$ and $< 10^{-9}$, respectively.  In both scenarios, the black, blue, and gray regions represent the bounds obtained for heavy neutrino lifetimes of $\tau_N = 1, 100, 1000$ ps, respectively. We also plot the exclusion limits obtained from searches on $|\Delta L|=2$ channels, $K^- \rightarrow \pi^{+}\mu^{-}\mu^{-}$ (NA48/2)~\cite{CERNNA48/2:2016} and $B^{-} \to \pi^{+}\mu^{-}\mu^{-}$ (LHCb)~\cite{LHCb:2014}, for comparison. For the $B^{-} \to \pi^{+}\mu^{-}\mu^{-}$ channel, we compare with the revised limit \cite{Shuve:2016} from the LHCb analysis~\cite{LHCb:2014}. 
The limit from $K^- \rightarrow \pi^{+}\mu^{-}\mu^{-}$ channel is taken for $\tau_N =$ 1000 ps~\cite{CERNNA48/2:2016}.
We can observe that the most restrictive constraint is given by $K^- \to \pi^+ \mu^-\mu^-$, which can reach $|V_{\mu N}|^2\sim \mathcal{O}(10^{-5})$, but only for a very narrow mass window of [0.25, 0.38] GeV. For $m_N > 0.38$ GeV, the CKM-suppressed four-body cha\-nnel $B_s^0 \to  K^- \pi^-\mu^+ \mu^+$ would complement the region of $|V_{\mu N}|^{2}$ covered by the channel $B^- \to \pi^+\mu^-\mu^-$ (also CKM-suppre\-ssed).

\begin{figure}[!t]
\centering
\includegraphics[scale=0.43]{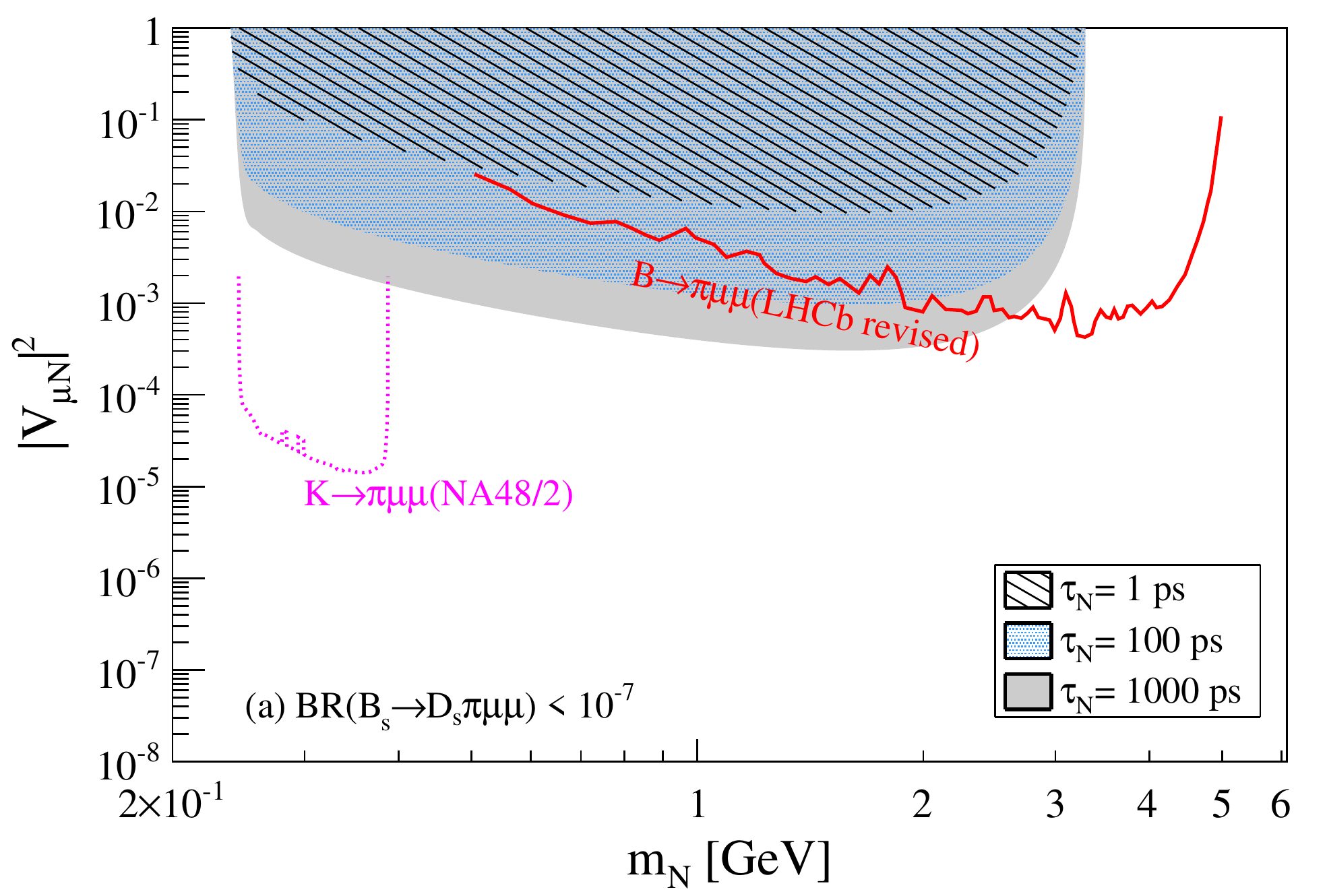} 
\includegraphics[scale=0.43]{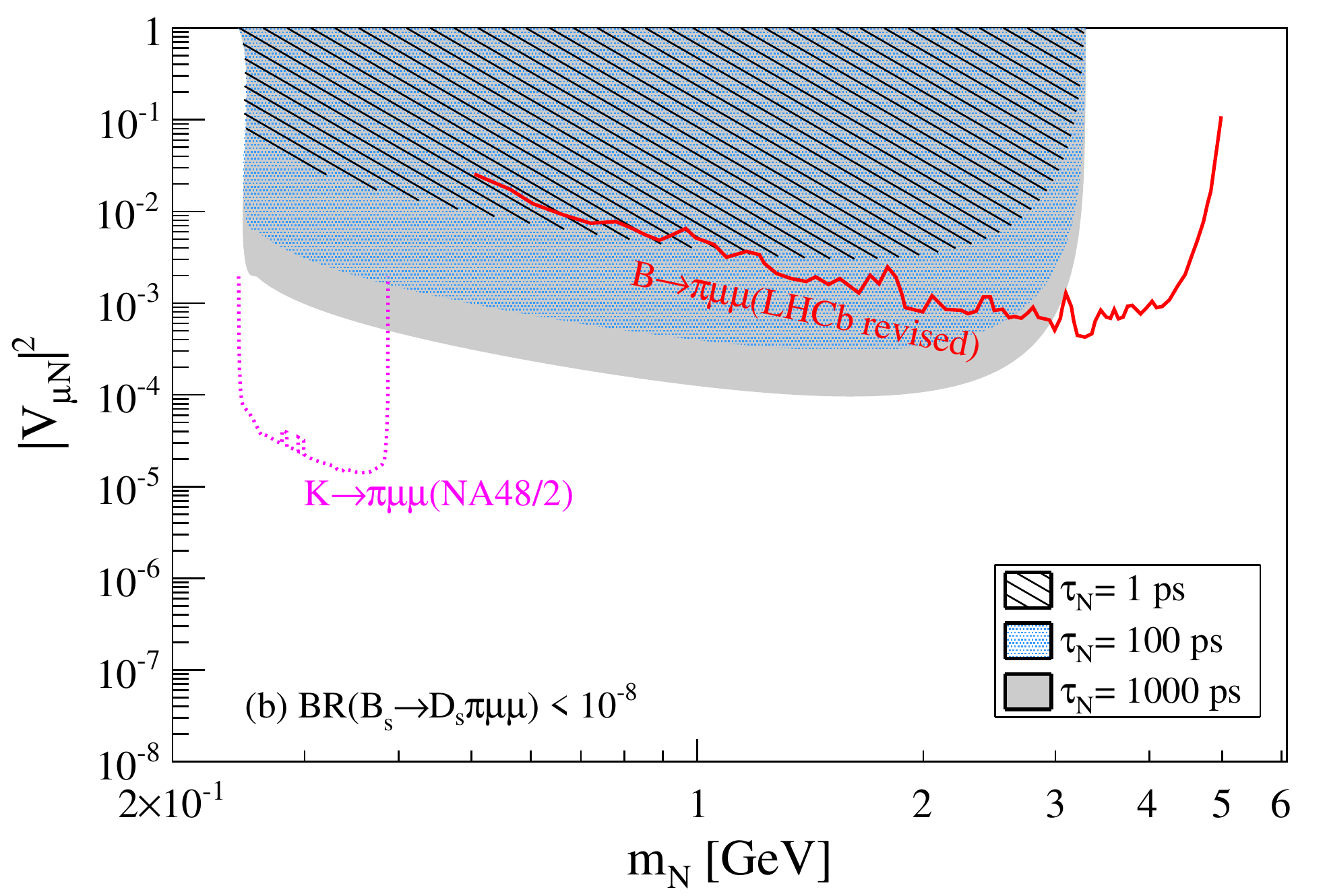}
\caption{\small The same caption as in Fig.~\ref{Fig4.1} but for (a) ${\rm BR}(B_s^0 \to  D_s^- \pi^-\mu^+ \mu^+) < 10^{-7}$ and  (b) ${\rm BR}(B_s^0 \to  D_s^- \pi^-\mu^+ \mu^+) < 10^{-8}$.}
\label{Fig4.2}
\end{figure}

For searches on $B_s^0 \to  D_s^- \pi^-\mu^+ \mu^+$, in Figs. \ref{Fig4.2}(a) and \ref{Fig4.2}(b), we plot the exclusion curves on the $(m_N,|V_{\mu N}|^2)$ plane for an expected sensitivities at the LHC of ${\rm BR}(B_s^0 \to  D_s^- \pi^-\mu^+ \mu^+) < 10^{-7}$ and $< 10^{-8}$, respectively. Again,  the black, blue, and gray regions represent the constraints obtained for heavy neutrino lifetimes of $\tau_N = 1, 100, 1000$ ps, respectively. For Majorana neutrino masses larger than 0.38 GeV, the $B_s^0 \to  D_s^- \pi^-\mu^+ \mu^+$ channel (CKM-allowed) would be able to exclude a slightly wider region of $|V_{\mu N}|^{2}$ than $B^- \to \pi^+\mu^-\mu^-$. The reason for this is the non-suppression for the CKM elements involved.

\begin{figure}[!t]
\centering
\includegraphics[scale=0.43]{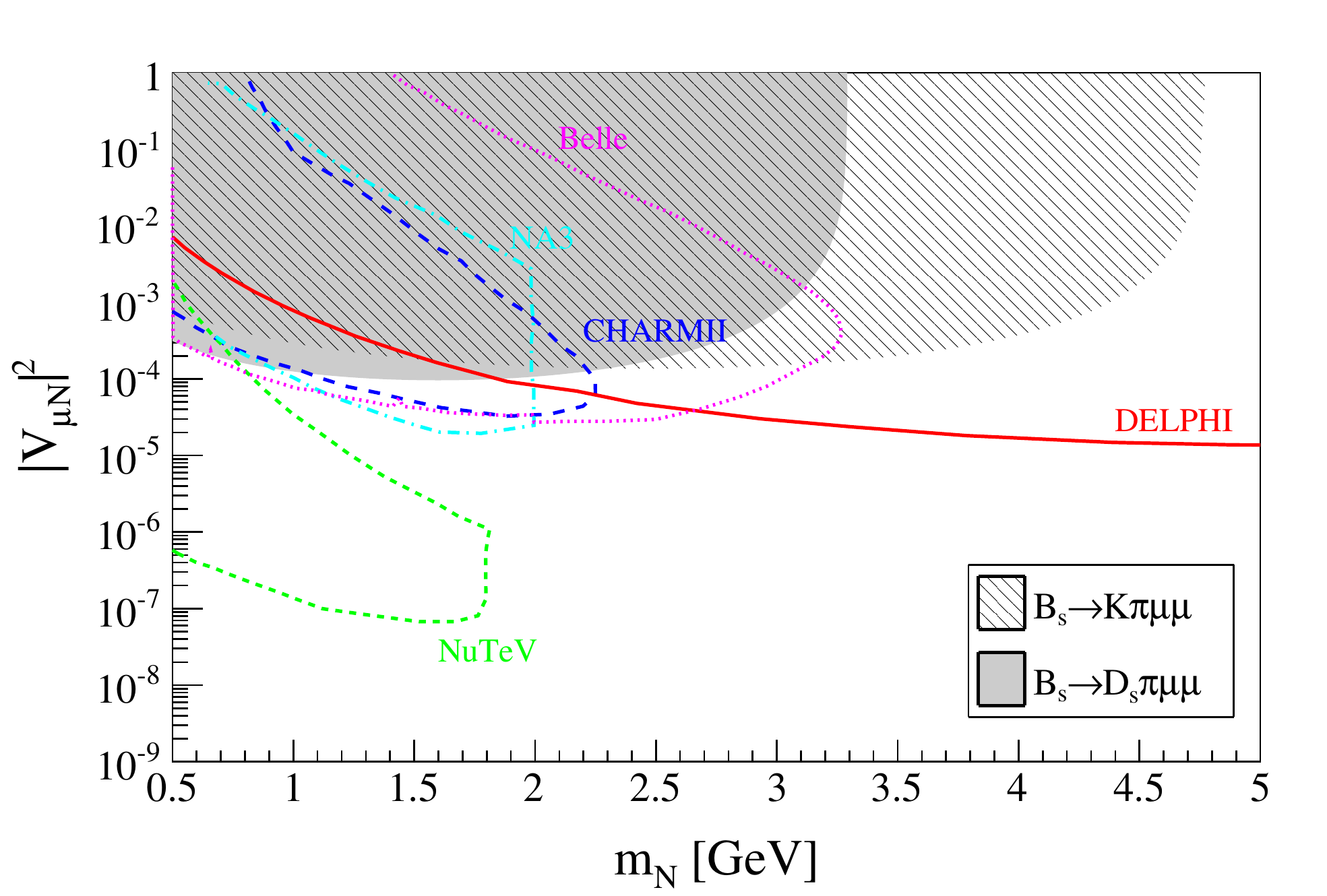}
\caption{\small Exclusion regions on the $(m_N, |V_{\mu N}|^{2})$ plane coming from the Belle \cite{Belle:N}, DELPHI \cite{LEP}, NA3 \cite{NA3}, CHARMII \cite{CHARMII}, and NuTeV \cite{NuTeV} experiments. Limits provided by the searches on $B_s^0 \to (K^-,D_s^-) \pi^-\mu^+ \mu^+$  are represented by the gray and black regions, respectively. See the text for details.}
\label{Fig4.3}
\end{figure}

Additionally, in Fig.~\ref{Fig4.3}, we show the exclusion bounds on the parameter space $(m_N,|V_{\mu N}|^2)$ coming from the Belle~\cite{Belle:N}, DELPHI~\cite{LEP}, NA3~\cite{NA3}, CHARMII~\cite{CHARMII}, and NuTeV~\cite{NuTeV} experiments, in the mass range [0.5,5.0] GeV\footnote{For recent reviews on the theoretical and experimental status of different GeV-scale heavy neutrino search strategies see Refs.~\cite{Atre:2009,Deppisch:2015,Drewes:2013,Drewes:2015,deGouvea:2015,Fernandez-Martinez:2016} and references therein.}. In comparison, the constraints obtained from the searches on $B_s^0 \to (K^-,D_s^-) \pi^-\mu^+ \mu^+$  are represented by the gray and black regions, for branching fractions of ${\rm BR} < 10^{-9}$ and ${\rm BR} < 10^{-8}$, respectively. In both cases, a lifetime of $\tau_N = 1000$ ps has been taken as a representative value. It is observed that our $|\Delta L|=2$ channels proposals are less restrictive than the bounds obtained from different search strategies, for instance, Belle~\cite{Belle:N} and DELPHI~\cite{LEP}. Nevertheless, keeping in mind that we have taken the most conservative values for the branching fractions derived in Secs.~\ref{LHCb} and~\ref{CMS}, it is possible that branching fractions values of the order  ${\rm BR} < 10^{-10}$ might be accessible to the LHCb and CMS experiments (see Figs.~\ref{Fig:1} and~\ref{Fig:2}), therefore, these $|\Delta L|=2$ channels would eventually provide complementary bounds.

\section{Concluding remarks} \label{Conclusion}

We have studied the semileptonic $|\Delta L|=2$ decays of the $B_s$ meson  $B_s^0 \to P^-\pi^-\mu^+\mu^+$ via the intermediate GeV-scale on-shell Majorana neutrino $N$, namely, $B_s^0 \to P^-  \mu^+N( \to \pi^-$ $\mu^+)$, with $P=K,D_s$. To our knowledge, these LNV decays of the $B_s$ meson have not been investigated before from a theoretical nor from an experimental point of view. We investigated these same-sign $\mu^+\mu^+$ channels and explored the sensitivity that can be reached at the LHCb and CMS experiments. We considered heavy neutrino lifetimes in the experimental (LHCb and CMS) accessible ranges of $\tau_N = [1, 100, 1000]$ ps, where the probability for the on-shell neutrino $N$ decay products to be inside the detector (acceptance factor $P_N$) has been taken into account in our analysis. As an outcome, it was found that for integrated luminosities collected of 10 and 50 fb${}^{-1}$ by the LHCb experiment and 30, 300, and 3000 fb${}^{-1}$ by the CMS experiment one would expect sensitivities on the branching fractions of the orders ${\rm BR}(B_s^0 \to K^- \pi^-\mu^+ \mu^+) \lesssim \mathcal{O}(10^{-9} - 10^{-8})$ and  ${\rm BR}(B_s^0 \to D_s^- \pi^-\mu^+ \mu^+) \lesssim \mathcal{O}(10^{-8} - 10^{-7})$, as conservative values. For masses in the ranges $m_N \in [0.25,4.77]$ GeV and $m_N \in [0.25,3.29]$ GeV, respectively, we extracted bounds on the parameter space $(m_N,|V_{\mu N}|^2)$ that might be obtained from their experimental search. Depending on the $\tau_N$ value, it was found that for $m_N > 0.38$ GeV these four-body channels may be capable of excluding a slightly wider region of $|V_{\mu N}|^{2}$ than $B^- \to \pi^+\mu^-\mu^-$ (LHCb). 

Consequently, the LHCb and CMS experiments have a great chance to look for heavy Majorana neutrinos in the near future, via $|\Delta L|=2$ decays of the $B_s$ meson. In addition, in the best-case scenario, the experimental search of these LNV channels would complement the bounds given by different search strategies (such as NA3, CHARMII, NuTeV, Belle, and DELPHI). 

\acknowledgments

The work of J. Mej\'{i}a-Guisao has been financially supported by Conacyt (M\'{e}xico) under Projects No. 254409, No. 221329 and No. 250607 (Ciencia B\'{a}sica), and No. 2015-2-1187 (Fronteras de  la  Ciencia). N. Quintero acknowledges support from Direcci\'{o}n General de Investigaciones - Universidad Santiago de Cali under Project No. 935-621717-016.  J. D. Ruiz-\'{A}lvarez gratefully acknowledges the support of COLCIENCIAS (Colombia).



\end{document}